\documentclass{aa}

\begin{document}

\thesaurus{    11.06.2  
               13.09.1 
	       11.19.6  
	       11.19.2  
	       11.19.5  
	       11.07.1   
}

\title{ Toward a dust penetrated classification of the evolved
stellar Population II disks
of galaxies }

\author{ David L. Block
\inst{1}
\and Iv\^anio Puerari
\inst{2} }
\institute{ Dept of Computational and Applied
Mathematics, University Witwatersrand, Private Bag 3,
WITS 2050, South Africa
\and Instituto Nacional de Astrof\'\i sica, Optica y Electr\'onica,
Calle Luis Enrique Erro 1, 72840 Tonantzintla, Puebla, M\'exico}

\offprints{ D.L. Block (igalaxy@iafrica.com)}
\date{\today}
\maketitle
\markboth{Classification of dust penetrated disks}{}
\begin{abstract}

To derive a coherent physical framework for the excitation
of
spiral structure in galaxies, one must consider the co-existence of
{\it two different} dynamical components: a gas-dominated Population
I disk (OB associations, HII regions, cold interstellar HI gas) and an
evolved stellar Population II component. The Hubble classification
scheme has as its focus, the morphology of the Population I component
only. In
the near-infrared, the morphology of evolved stellar disks indicates a
simple classification scheme: the dominant Fourier m-mode in the dust
penetrated regime,
and the associated pitch angle.
On the basis of deprojected K$'$ (2.1$\mu m$) images, we
propose that the
evolved stellar disks may be grouped into three principal
dust penetrated archetypes:
those with tightly wound stellar arms characterised by pitch angles at
K$'$ of $\sim$ 10$^{\circ}$ (the $\alpha$ class),
an intermediate group with pitch angles of $\sim$ 25$^{\circ}$ (the
$\beta$ class) and
thirdly, those with open spirals demarcated by pitch angles at K$'$ of
$\sim$ 40$^{\circ}$ (the $\gamma$ bin).

There is no correlation between our dust penetrated classes and
optical Hubble binning; the Hubble tuning fork
does not constrain the morphology of the old stellar Population II
disks.
Any specific dust penetrated archetype may be the resident
disk of {\it both} an early or late type galaxy.
The number of arms and the pitch angle of the arms
at K$'$
of the early-type `a' spiral NGC 718 are almost identical to those for the
late-type `c' spiral NGC 309.
We demonstrate that galaxies on opposite
ends of the tuning fork can display remarkably similar evolved disk
morphologies and belong to the same
dust
penetrated class. Furthermore, a prototypically flocculent galaxy such
as NGC 5055 (Elmegreen arm class 3) can
have an evolved disk morphology almost identical to that of NGC 5861,
characterised in the optical as having one of the most regular
spiral patterns known and of Elmegreen class 12.
Both optically flocculent or grand design galaxies can reside
within the {\it same} dust penetrated morphological bin.
As was suggested by Block et al. (\cite{block94a}), it is the gas
dominated Population I component
which determines the optical types (a, b, c), decoupled from the
Population II.

Those L=lopsided galaxies (where m=1 is a dominant mode) are
designated
L$\alpha$, L$\beta$ and L$\gamma$ according to the dust penetrated pitch
angle; E=evensided galaxies (where m=2 is the dominant
Fourier mode) are classified into classes E$\alpha$, E$\beta$ and
E$\gamma$, according to our three principal dust penetrated archetypes.
The L and E modes are the most common morphologies in our sample,
which spans a range
of Hubble types from early (a) to late (irregular). Having formulated
our dust penetrated classification
scheme here, we have tested it on an independent sample of 45 face-on
galaxies observed in the near-infrared by Seigar and James
(\cite{seigar98a}, b).


\end{abstract}


\section{Introduction}

There was a time when there appeared to be very little interconnection
between the presence and evolution of dust
grains on the one hand, and galaxy morphology on the other.
This has now changed so dramatically, that an entire International
Conference was recently devoted to the {\it interrelation} between these two
apparently discrepant disciplines (see Block
\cite{block96a}).

Allen (\cite{allen96}) summed it
up thus:

{\it ``We're now looking at a transition to a possible change in the
way we look
at galaxies.  Sometimes ... we see disks that have a spiral structure that
we couldn't have dreamt existed from looking at the optical picture  ...
we've got a possibility here of applying the morphology to a physical
framework, perhaps in a way that none of us could have dreamt of before we
had the capability of sweeping the dust away from the galaxy in a figurative
sense''}.

Optically thick dusty domains in galactic disks can completely obscure
underlying stellar structures. The presence of dust and the morphology of a
galaxy are inextricably intertwined: indeed, the morphology of a galaxy can
completely change once the disks of galaxies are dust penetrated (eg.
Block
and Wainscoat \cite{block91}; Block, Elmegreen and Wainscoat
\cite{block96b}).

The distribution of dust grains in the bulges and disks of galaxies may be
very widespread (Block et al. \cite{block94b}), even extending to the outer
parts of galaxies (Lequeux and Gu\'elin \cite{lequeux96}). The tracing
of dust grains in nature can be very elusive:
high levels of dust extinction do {\it not} necessarily imply that the
effects of dust attenuation (ie. observed reductions in surface brightness
profiles) will also be large, because scattering by dust grains may fill in
at least part
of the lost surface brightness. The effective albedo of dust grains in the
near-infrared can actually be higher than in the optical (for a
determination of the near-infrared albedo of dust grains in M51, see Fig. 4
in Block \cite{block96a}).

Furthermore, large amounts of dust do not necessarily imply red V-K
(K: 2.2$\mu m$) colours. The dust column density on the far side of an
inclined spiral
with relatively {\it blue} V-K colours can be just as high as on the near
side, where the V-K colour distribution can be much redder (Elmegreen and
Block \cite{elmegreen98}).

The classification of galaxies has traditionally been inferred from
photographs/CCD imaging shortward of the 1 micron window, where
stellar Population II disks are not yet dust penetrated. The NICMOS and
other near-infrared camera arrays offer unparralleled opportunities for
deconvolving the Population I and II morphologies, because the opacity at
K -- be it due to absorption or scattering
-- is always low. The extinction (absorption+scattering) optical depth at K
is only 10\% of that in the V-band (Martin and Whittet \cite{martin90}).

From a dynamical viewpoint, the disk of a spiral galaxy can be
separated into two distinct components: the {\it gas--dominated} Population I
disk, and the {\it star--dominated} Population II
disk. The former component contains features of spiral structure
(OB associations, HII regions,
and cold interstellar HI gas), which are naturally fast evolving; dynamically,
it is very active and responsive, because, being
characterized by small random motions (a ``cool disk''), it fuels Jeans
instability. In contrast, the Population II disk, which is dynamically
``warmer'' because of the larger epicycles, contains the old
stellar population betraying the underlying mass distribution -
it is the `backbone' of the galaxy
(Lin \cite{lin71}).
One might expect --
even in the absence of
appreciable optical depths -- for the two morphologies to be
very different, since the
near-infrared
light comes predominantly from giant and supergiant stars (Rix and Rieke
\cite{rix93b}, Frogel et al. \cite{frogel96}).

It is important to stress that from this physical (dynamical) point of
view, one therefore requires {\it two} classification
schemes -- one for the Population I disk, and a separate one for
the
Population II disk.
A near-infrared classification scheme can
never {\it replace}
an optical one, and vice-versa, because the {\it current}
distribution of old stars strongly affects the {\it current}
distribution of gas in the Population I disk.
The dynamic interplay between the two
components -- via a {\it feedback} mechanism -- is crucial, and has been
studied extensively (eg. Bertin and Lin \cite{bertin96b}, who term
this mechanism a dynamical thermostat).

A central aspect here is the {\it likely coupling}
of the Population I disk with that of the Population II disk via
the feedback mechanism. To
quote Pfenniger et al. \cite{pfenniger96}, {\it ``The interesting aspect
of this coupling is that the systematic global properties of galaxies
are then no longer necessarily determined by the initial conditions of
collapse.''}

We find a
duality
in spiral structure. A typical turbulent speed associated with the cold
gas Population I component is $\sim$ 6 km s$^{-1}$; in contrast, the
velocity dispersion for old stars would typically be six times larger.
While a gas cloud may be constrained to move in a thin annulus $\Delta r
\sim 300\,\,pc$, an old star may wander in an annulus 2 kpc thick
(Bertin and Lin \cite{bertin96b}).
To derive a {\it coherent} physical framework for the excitation
of
spiral structure in galaxies, one must consider the co-existence of the
{\it two} dynamical components.

{\it There is a
fundamental limit
in predicting what evolved stellar disks might
look like}. The greater the degree of decoupling, the greater is the
uncertainty.
The fact that a spiral might be
flocculent in the optical
is very important, but it is equally important to know whether or not
there is a  decoupling with a
grand design old stellar disk.
No prediction on that issue can, a priori, be made (Block, Elmegreen and
Wainscoat \cite {block96b}).

The theoretical framework to explain the co-existence of
completely
different morphologies within the same galaxy when it is studied
optically and in the near-infrared is beautifully described by Bertin
and Lin (\cite{bertin96b}), following the early pioneering work of
Lindblad (\cite{lindblad63}). A global mode (Bertin et al.
\cite{bertin89a}, \cite{bertin89b}) can be imagined as being composed of
spiral wavetrains propagating radially in opposite directions, much like
a standing wave. Thus a feedback of wavetrains is required
from the center. The return of wavetrains
back to the corotation circle is guaranteed by refraction, either
by the bulge or because the inner disk is
dynamically warmer. In the {\it stellar} disk, such a feedback can be
interrupted by Inner Lindblad Resonance (ILR),
which is a location where the stars meet the
slower rotating density wave crests in resonance with their epicyclic
frequency (Mark \cite{mark71}; Lynden-Bell and Kalnajs
\cite{lynden72}). In the {\it gaseous} disk,
the related resonant absorption is only partial, so that some feedback
is guaranteed. Once the above described {\it wavecycle} is set up (in the
absence of a cutoff by ILR), a self--excited global mode can be generated.

The tightness of the arms in the modal theory
comes from the mass distribution and rate of shear.
Galaxies with more mass concentration, i.e., higher
overall densities (including dark matter)
and
higher shear, should have more tightly wound arms.

If the disk is very light -- low $\sigma$ where $\sigma$ is the disk
density -- the mode can be very tight, and one is in the domain of
small epicycles (formally,
the stability parameter Q = $c\kappa/\pi G \sigma $
being close to unity, the value of $c$ must also be small. Here $c$ is the
radial velocity dispersion
and $\kappa$ is the epicyclic frequency).

If one increases the mass of the disk one finds a trend toward more
open
structures, but soon one runs the risk of a disk that is too heavy
and a bar
mode should manifest itself.

In those spirals with more open spiral arms at
K but with no sign of a prominent bar, Bertin
(\cite{bertin96a}) anticipates that
the galaxy should be {\it gas rich}. Abundant gas can shock, dissipate
and
make some violently unstable {\it open} modes (see frame d of Fig. 4.5
in the book by Bertin and Lin \cite{bertin96b}).
This is one important reason why Bertin and Lin place `gas content' in
their
theoretical framework for classification as the governing parameter for
the
trend from `a' to `c' galaxies (as indicated in Fig. 1 of Block et al.
\cite{block94a}).

The redistribution of angular momentum by large-scale spiral torques
will be stronger for stellar arms which are more open; some authors
(eg. Pfenniger et al. \cite{pfenniger96}) have
postulated that such a redistribution may lead to rapid changes in the
disk and even modify the properties of the rotation curve.
This is the concept of secular evolution of a galaxy,
from an open to a more tightly wound morphology, within one Hubble time.

Many galaxies show the presence of a significant m=1
component
in the near-infrared (often in the form of a lopsidedness of the
spiral). The
linear modal theory predicts that m=1 modes
should generally be dominated by m=2 modes when available, since the
latter are more efficient in transporting angular momentum
outwards.
However, modes greater than m=2 are generally suppressed
in the stellar disk by ILR (but see the discussion in section 7).
While the disk mass
participating in the mode is crucial, the gas-content of the
galaxy is important: gas-rich spirals can generate modes greater than
m=2. It had earlier been predicted (Block et al. \cite{block94a},
Bertin and Lin \cite{bertin96b}) that infrared images should show
{\it an ubiquity
of global one and two armed structures in the underlying stellar disk}
and we believe that any classification of dust penetrated disks at K
should
indicate this number.

In contrast,
the dynamics of the cold Population I gaseous disk, characterized
by different scalelengths, velocity dispersion, thickness, and behaviour at
the relevant Lindblad resonances, explains why
spiral galaxies are optically so often overwhelmed by higher m modes and
other more irregular fast evolving features,
supported by the cold interstellar gas (Bertin \cite{bertin91},
\cite{bertin93}).

\section{The decomposition and identification of modes}

Our sample in Table 1 was selected
to cover as wide a range of optical Hubble types as possible -- from
early type to irregular. It spans a range in blue absolute magnitude
from $-19.7$ to $-23.3$,  a range in linear
diameter from 14 kpc to 97 kpc (the galaxies at the lower and upper
limits of these ranges are NGC 1637 and NGC 309) and a wide range in
linear arm width. From a sample of several hundred galaxies which one of
us (DLB) has studied on
a uniform linear scale, NGC 309 has some of the
linearly widest spiral arms known (see Figure 1 in Block
(\cite{block82})).

The near-infrared imaging of many of the spirals
was secured at Mauna Kea and in Chile and fully described in Block et
al. (\cite{block94a}).
At Mauna Kea, a Rockwell HgCdTe 256 $\times$ 256 camera array was used,
sensitive from
1 to 2.5$\mu m$. At
the f:10 focus of the University of Hawaii's 2.2m telescope, with 1:1
re--imaging optics, the scale is 0.37$''$/pixel, giving a
field of 95$''$ $\times$ 95$''$.
At the European Southern Observatory at La Silla, we used the IRAC2
camera mounted on the ESO 2.2m telescope, in mode C
(0.49$''$/pixel, giving a 2$'\times2'$ field of view).
Both at La Silla and Mauna Kea, the K$'$ filter (2.1$\mu m$) was always
selected.
This filter is similar to K, but shifted 0.1 $\mu m$ towards the blue,
to reduce the thermal background (Wainscoat and Cowie
\cite{wains92}).

The K$'$ images of additional galaxies were very kindly made
available to us by Dr P. Grosb{\o}l (NGC 3223, NGC 5085, NGC
5247, NGC 5861 and NGC 7083), Dr M. Verheijen (NGC 3893, NGC 3938,
NGC 3992 and NGC 4051) and Dr M. Thornley (NGC 3521 and NGC 5055).

The 2-D Fast Fourier decomposition of all the K$'$ images in
this study, employed a program developed
by Puerari (Schr\"oder et al. \cite{schroder94}). In the
decomposition, logarithmic spirals of the form
$r=r_o {\rm exp} (-m \theta /p_{\rm max})$
are employed.

The amplitude of each Fourier component is given by (Schr\"oder et al.
\cite{schroder94})

$$\displaystyle A(m,p) = \frac{\Sigma_{i=1}^I \Sigma_{j=1}^J I_{ij}
({\rm ln}\;r,\theta)\; {\rm exp}\;(-i(m \theta + p\; {\rm
ln}\;r))}{\Sigma_{i=1}^I \Sigma_{j=1}^J I_{ij} ({\rm ln}\;r,\theta)}$$

\noindent where $r$ and $\theta$ are polar coordinates, $I({\rm
ln}\;r,\theta)$
is the intensity at position $({\rm ln}\;r, \theta)$, $m$ represents the
number of
arms or modes, and $p$ is the variable associated with the pitch angle
$P$,
defined by $\displaystyle \tan P = -\frac{m}{p_{\rm max}}$.

\begin{figure}
     \vspace{19cm}
       \includegraphics{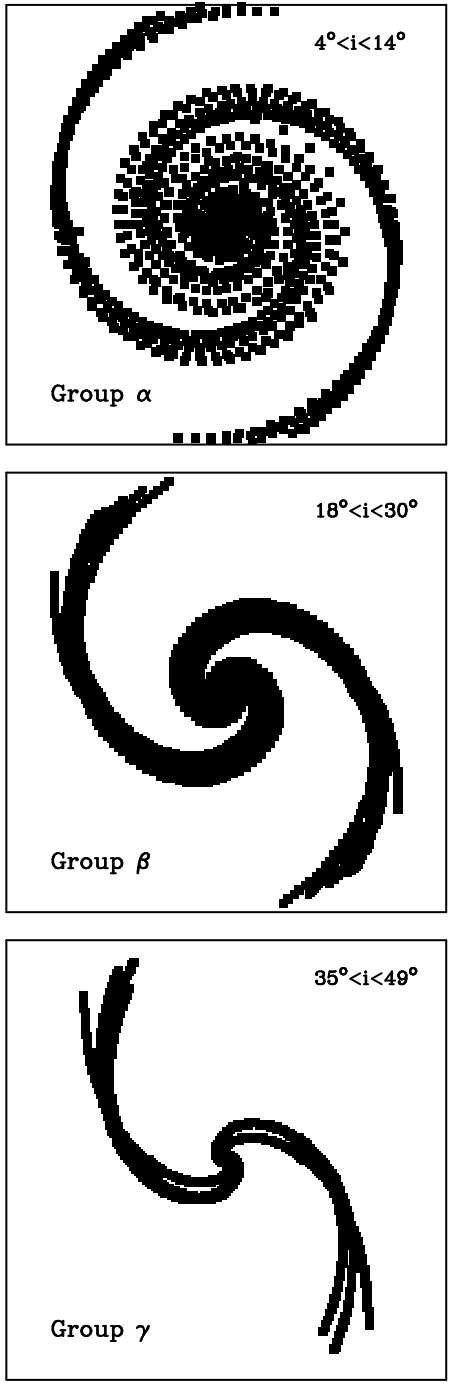}
    \caption{Logarithmic spirals with pitch angles determined
from the Fourier spectra suggest three principal archetypes $\alpha$,
$\beta$ and $\gamma$ for the classification of dust penetrated
evolved Population II disks. Individual pitch angles are listed in
Table 1.}
\end{figure}

Our analytic Fourier spectra corroborate earlier observational
indications (Block et al. \cite{block94a})
that there is indeed an ubiquity of m=1 and m=2 modes, and that three
principal
archetypes for the evolved stellar disk of such galaxies may be
proposed, based on the the pitch angle of the arms at K$'$ (see Figure
1). Figure 1 is generated from the actual deprojected images at K$'$ of
our galaxies where the pitch angles have robustly been
determined from the Fourier spectra. It is
apparent that a galaxy
with a winding angle at K$'$ of 18 degrees looks remarkably similar to
one with a winding angle of 30 degrees (see Figure 1), and we classify
those into one distinct bin.
We choose to designate our three dust penetrated (DP)
bins as $\alpha$, $\beta$ and $\gamma$, depending on the pitch angle of
the dominant m-spiral at K$'$.

Those L=lopsided galaxies (where m=1 is a dominant mode) are
designated
L$\alpha$, L$\beta$ and L$\gamma$ according to the dust penetrated pitch
angle; E=evensided galaxies (where m=2 is the dominant
Fourier mode) are classified into classes E$\alpha$,
E$\beta$ and
E$\gamma$, according to our three principal dust penetrated archetypes
(see column 2 of Table 1). It is proposed that higher order modes (which
may exist under
special circumstances: see section 7 of this paper) be classified as H3
(for m=3) and H4 (for m=4), followed by a class description $\alpha$,
$\beta$ or $\gamma$ appropriate to the pitch angle of the stellar arms.

\begin{figure}
     \vspace{10cm}
       \includegraphics{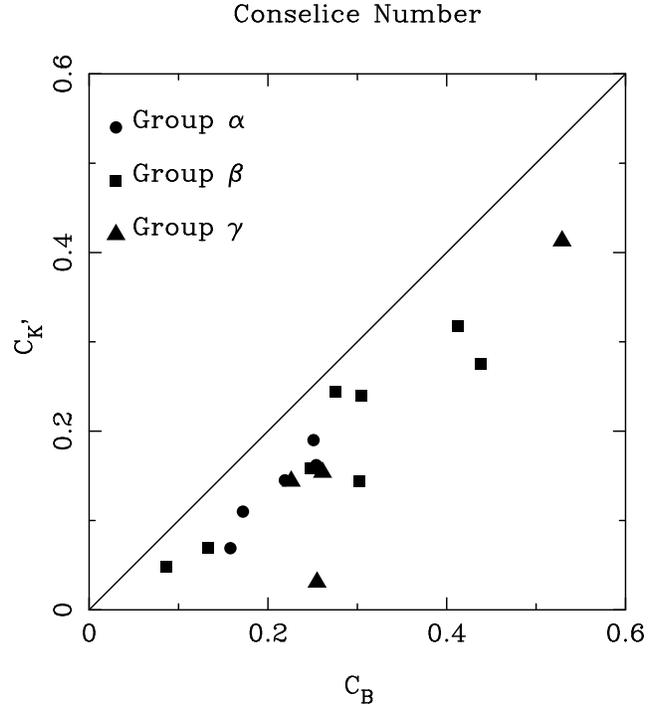}
       \caption{The Conselice symmetry number confirms that the disks
of spiral galaxies are much more regular in the near-infrared,
and therefore
have a lower value at K$'$ compared to the value in the optical (B)
regime.}
\end{figure}

The terminology DP
highlights the fact that at K$'$, one is looking at
dust penetrated  images. It is dust in the gaseous
Population I disk which so often obscures the underlying Population
II disk. It is because of dust that a galaxy can masquerade
as late type c (for the optical morphology) but as early type a in
the
dust penetrated Population II regime (Block and Wainscoat
\cite{block91}).
It is dust which  obscures the regular two-armed spiral structure
found in many optically flocculent galaxies
(eg. Grosb{\o}l and Patsis
\cite{grosbol96}, \cite{grosbol98}, Thornley \cite{thornley96}). Optical
images, combined with near-infrared ones,
together with the use of radiative transfer models highlight the fact
that gaseous Population I disks often have optical depths $\tau(V)
\sim 1$ and more (Block et al. \cite{block94b}, Elmegreen and Block
\cite{elmegreen98}) which
is why there can be no {\it prediction} of what an evolved stellar disk
of a optically flocculent galaxy, for example, might
look like
(Block, Elmegreen
and Wainscoat \cite{block96b}).

We
have avoided using I, II or III for our dust penetrated classes
and have used $\alpha$, $\beta$ and $\gamma$ instead, to prevent
confusion with the optical luminosity classification scheme of galaxies
(van den Bergh \cite{bergh60}).

While there may appear to be a resemblance
of our groups to the Hubble bins in terms of the openness or pitch
angle of the arms, there
is no similarity whatever. It is possible for a late type Sc galaxy (eg.
NGC 5861 below) to have
a {\it stellar disk with tightly wound arms} with a pitch angle at
K$'$ of 13$^{\circ}$, which demarcates it to our tightly wound
group $\alpha$
class. Conversely, it is possible for  two galaxies on opposite
ends of the Hubble sequence (a, c)
to {\it both} have evolved disks of dust penetrated group $\beta$: NGC
718 and NGC 2997 are examples which both belong to group $\beta$, but
have
Hubble classifications of a and c respectively. (The near-infrared pitch
angles of NGC 718 and NGC 2997 differ by only
1$^{\circ}$ -- see Table 1.) It is also possible
for an optical irregular to have smooth stellar spiral arms with a
well-defined pitch angle at K$'$ and to belong
to one of these groups (eg. the companion to M51, as discussed below).

\begin{table}
\caption[]{Optical Hubble type, dust penetrated (DP) morphologies,
near-infrared pitch angles and radial range of fit.}

\hoffset-2.0truecm

\begin{tabular}{cccccc}

GALAXY  &  DP Class & Morph. Type   &   {\it P}$_{\rm K'}$
& Range of fit \\

\vspace{.3cm}

{\it Group $\alpha$}\\

NGC 2841  &  E$\alpha$ & Sb  & 9$^{\circ}$ & 70$''$, 165$''$ \\

NGC 3223 & E$\alpha$ & Sb & 9$^{\circ}$ & 15$''$, 48$''$ \\

NGC 3521 & E$\alpha$ & Sbc & 14$^{\circ}$ & 12$''$, 157$''$\\

NGC 3992  & E$\alpha$ &    SBbc     &   11$^{\circ}$  & 61$''$,
184$''$\\

NGC 4622 & E$\alpha$/L$\alpha$ & Sb & 8$^{\circ}$ & 11$''$, 50$''$\\

NGC 5055 & E$\alpha$ & Sbc & 13$^{\circ}$ & 18$''$, 150$''$ \\

NGC 5861 & E$\alpha$ & Sc & 13$^{\circ}$ & 17$''$, 50$''$ \\

\vspace{.3cm}

{\it Group $\beta$}\\

NGC 309 &  E$\beta$ &    Sc  &  18$^{\circ}$  & 11$''$, 44$''$ \\

NGC 718 &  E$\beta$ &    Sa       & 24$^{\circ}$  & 11$''$, 34$''$\\

NGC 1637 & E$\beta$/L$\gamma$ & Sc & 30$^{\circ}$ & 13$''$, 45$''$ \\

NGC 2997  &  E$\beta$ &  Sc       &   25$^{\circ}$ & 34$''$, 122$''$ \\

NGC 3893  & E$\beta$ &   Sc       & 21$^{\circ}$  & 12$''$, 88$''$\\

NGC 3938  &  E$\beta$ &  Sc       & 22$^{\circ}$  & 12$''$, 168$''$ \\

NGC 4736  & E$\beta$ & Sab & 28$^{\circ}$ & 37$''$, 56$''$ \\

NGC 5247 & E$\beta$ & Sc & 30$^{\circ}$ & 30$''$, 90$''$ \\

\vspace{0.3cm}

{\it Group $\gamma$}\\

NGC 4051  &  E$\gamma$ &  SBbc     & 49$^{\circ}$ & 16$''$, 180$''$ \\

NGC 5085 & E$\gamma$ & Sc &  39$^{\circ}$ & 25$''$, 50$''$ \\

NGC 5195  & E$\gamma$ &   Irr      &  49$^{\circ}$  & 26$''$, 52$''$
\\

NGC 7083 & E$\gamma$ & Sb & 36$^{\circ}$ & 15$''$, 40$''$ \\

\end{tabular}

\end{table}

The range of radii over which the fit was applied are all listed in
Table 1. These were selected to exclude the bulge (where there is no
information on the arms) and extend to the outer limits of the arms
in our images. It is important to discuss
the error bars on the determination of our pitch angles using Fourier
transform techniques. We use a Fast
Fourier Transform, where the incremental step-size in the spectrum is
$\Delta p$. We calculate the spectrum from $-50<p<50$ and select a
step-size of $\Delta p$=0.25.

The error bars decrease with the degree of tightness of the arm pattern.
In NGC 5195 for example -- with very open stellar arms -- the spectrum
peak occurs at $p$=1.75 (see Figure 7). We therefore compute a pitch
angle in Table
1 of $P$=atan($m$=2/$p$=1.75)=48.8$^{\circ}$. If the peak was to occur
one
step-size later, at $p$=2.0, the computed pitch angle would have been
$P$=45.0$^{\circ}$. If the Fourier spectrum was to have peaked at
$p$=1.5,
one step-size earlier, then $P$=53.1$^{\circ}$. So, for this galaxy, our
pitch angle at K$'$ of $P$=48.8$^{\circ}$ has an error bar of ($+4.3,-3.8$) degrees.

On the other hand, let us consider NGC 3992, with its tightly wound
arms. The Fourier spectrum peaks at $p$=10.25 (see Figure 3), and so the
derived pitch
angle in Table 1 is $P$=11.0$^{\circ}$. If the peak were to have
occurred
at $p$=10.0, the pitch angle would have been $P$=11.3$^{\circ}$. If the
spectrum peaked one step-size later at $p$=10.5, the derived pitch
angle would be $P$=10.8$^{\circ}$. In this case, the error bar for our
listed 11$^{\circ}$ tightly wound arm structure in NGC 3992 is ($+0.3,-0.2$) degrees.

In the tightly wound scenario, it is perhaps not too meaningful to quote
error bars on the pitch angle of only $\sim \pm 0.2^{\circ}$.
Spirals
with $P$=11.0$^{\circ}$ or $P$=11.2$^{\circ}$ will only be `different'
after
a number of winding angles or turns, and we know it is difficult to find
arms at K$'$ which actually wind round more than 360$^{\circ}$, although
such galaxies do exist (eg. Fig 6b in Block et al. \cite{block94a}).

Pitch angles can either be determined from the peaks in the Fourier
spectra (as above), or from the slopes in $({\rm ln}(r), \theta)$ plots,
where logarithmic
spirals appear as straight lines. In both cases, careful
deprojections to face-on are critical. Our experience with estimating
pitch angles from $({\rm ln}(r), \theta)$ plots is that eyeball
identification
of the straight lines (arms) and of their resulting slopes can be very
difficult,
especially if the arm/interarm contrast in the near-infrared image is
not high. On the other hand, peaks in Fourier spectra are unambiguous.

As other authors have advocated in the past (eg. Consid\`ere and
Athanassoula \cite{considere88}; Garc\'\i a-G\'omez and Athanassoula
\cite{garcia93}), the Fourier transform is the most powerful method to
find periodicity in a distribution, whether it be one dimensional
(such as
radial stellar pulsations) or bi-dimensional (as for the spiral arms in
an image). That there are galaxies with very open arm morphologies, with
pitch angles as large as $\sim$ 40-45$^{\circ}$, is corroborated by
these earlier studies.
For example, Consid\`ere and Athanassoula (\cite{considere88}) analysed
blue (B)
images of 16 galaxies, and found pitch angles as large as 37$^{\circ}$.
Garc\'\i a-G\'omez and Athanassoula (\cite{garcia93}) analysed the
distribution of HII
regions of 44 galaxies and also found some very open morphologies. For
the dominant m=2 component, Garc\'\i a-G\'omez and Athanassoula
(\cite{garcia93})
determine pitch angles of 43$^{\circ}$ for NGC 5194 and 55$^{\circ}$
for both NGC 3627 and NGC 7741. Many other examples are provided in
their table. These large pitch angles were only for the main m=2
component;
Garc\'\i a-G\'omez and Athanassoula (\cite{garcia93}) report even
larger pitch angles for other Fourier components which they see in their
spectra.

We have also computed the symmetry number proposed
by Conselice (\cite{conselice97}) for each of our B and K$'$ images:

$$\displaystyle C^2\equiv \frac{\Sigma \frac{1}{2} (I_0-I_{180})^2}
{\Sigma I_0^2}$$

\noindent The number is determined
by rotating
every image through 180$^{\circ}$ about the centre, subtracting that
from the original image and normalising.
We did this
to illustrate that highly
symmetric evolved disks exist for all three of our dust penetrated
groups
(see Figure 2). A value of $C_{\rm \lambda}$=0
corresponds to
a completely symmetric galaxy at wavelength $\lambda$; for a completely
asymmetric galaxy, the ratio is unity.

In contrast
to the
work of Conselice, where a homogeneous sample was used, our sample here
is drawn from
a non-homogeneous one (different instruments and different observers).
Care must be exercised when
working with images from different instruments: the potential problem is
that of noise -- especially in the near-infrared, where disk features
might be of a low contrast and where the sky is bright.
A
greater dispersion can be expected, but the quantitative trend observed
in Figure 2 confirms our earlier qualitative findings in Block et al.
(\cite{block94a}), that disks
at K$'$ are much more regular and symmetric than seen at shorter
wavelengths, such as B.

The Conselice number does not increase with dust penetrated class. For
example, its
value for NGC 5195 is only 0.031 at K$'$, confirming the
visual impression of a very symmetric evolved disk (see
Fig. 5b in Block et el. \cite{block94a}). While the
Conselice method might be used to
confirm the presence of symmetric disks in all three of our dust
penetrated archetypes, it is the
pitch angle of the arms at K$'$-- and not the Conselice number -- which
determines to which dust penetrated bin a galaxy should be assigned.

\section{Dust penetrated group $\alpha$}

Our Fourier spectra of these galaxies, with tightly wound stellar arms
at K$'$, are illustrated in Figures 3 and 4.

NGC 3223  is an optically multi-armed specimen of Hubble
type Sb and van den Bergh (\cite{bergh60}) luminosity class I-II.
In
the near-infrared, however, this galaxy shows two bright, {\it grand
design} stellar density waves (SDWs) (see Grosb{\o}l and Patsis
\cite{grosbol96}, \cite{grosbol98}).

\begin{figure*}
     \vspace{19.5cm}
       \includegraphics{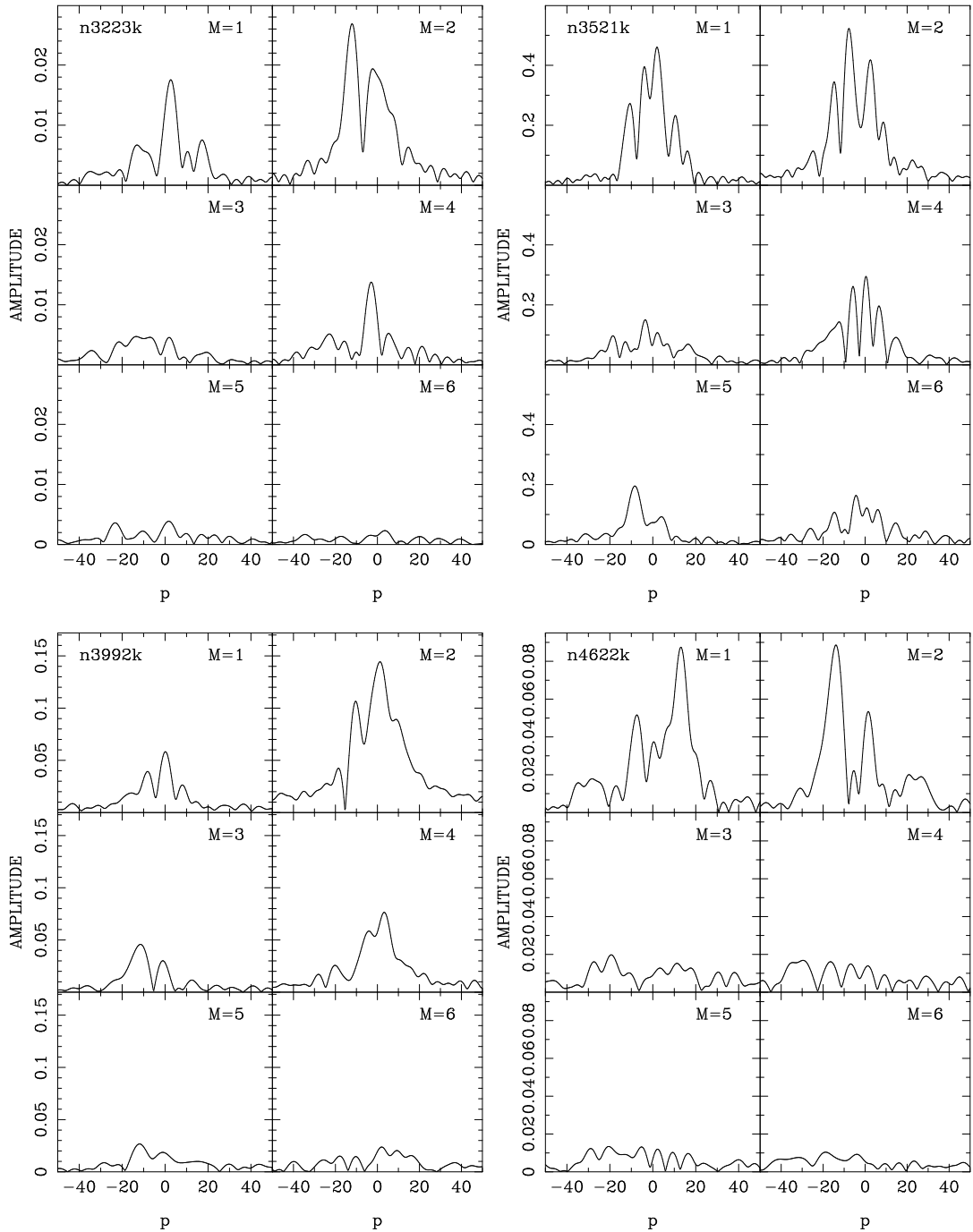}
       \caption{The Fourier spectra for our group $\alpha$ galaxies.
Although the dominant mode is m=2, note that some galaxies (eg. NGC
3233, NGC 3521 and NGC 3992) have significant higher order [m=4] modes
at K$'$. The single leading arm of NGC 4622 has a striking m=1 component
at K$'$.}
\end{figure*}

\begin{figure*}
     \vspace{10cm}
       \includegraphics{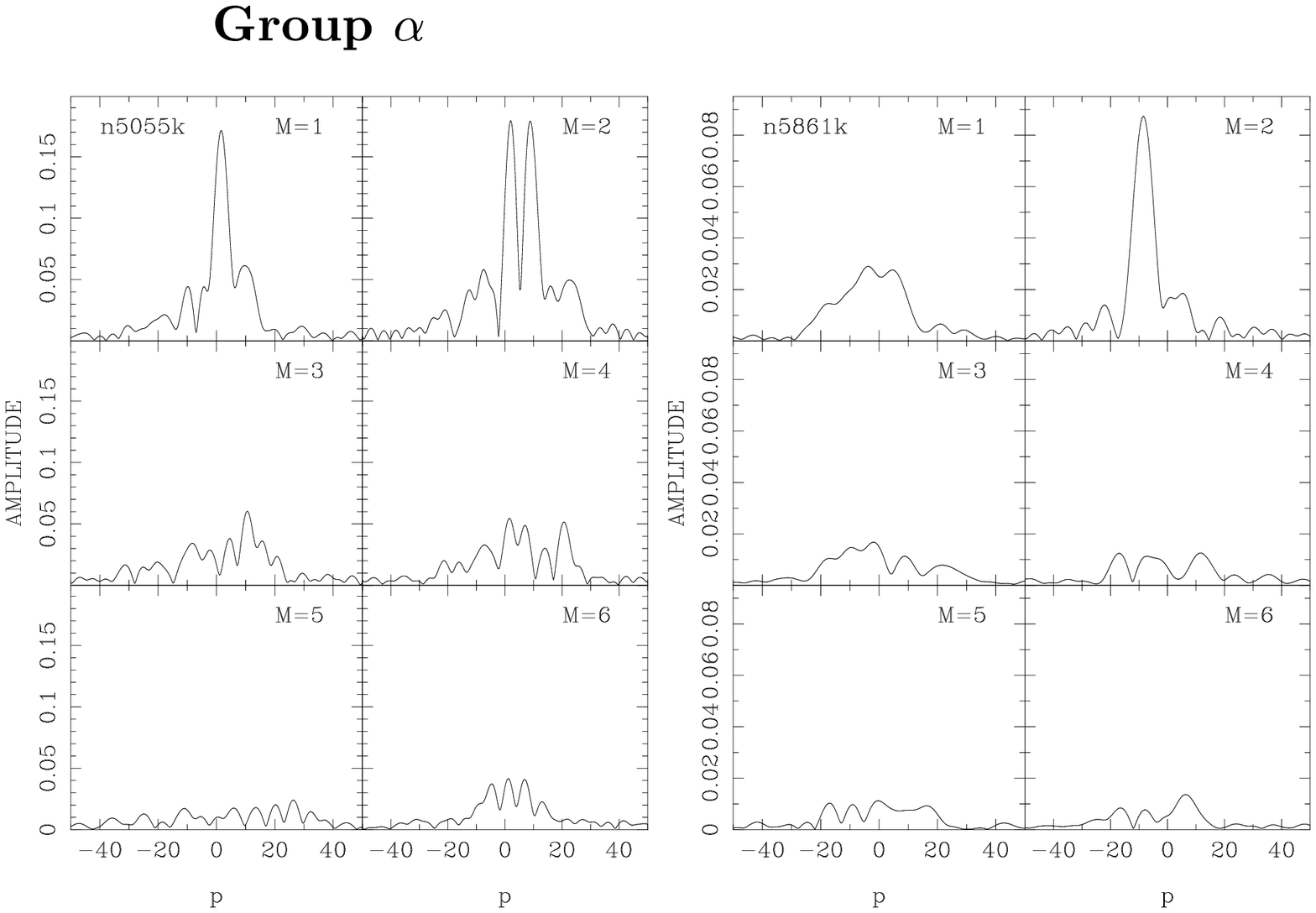}
       \caption{The Fourier spectra of additional galaxies in our
dust penetrated $\alpha$ group shows an ubiquity of m=1 and m=2 modes.}
%
\end{figure*}

That multi-arm galaxies such as NGC 3223 may present long, continuous
arms with well-defined pitch angles at K$'$ is not uncommon: for the
famous optically flocculent prototype NGC
2841 (Elmegreen arm class 3 = `fragmented arms uniformly distributed
around
the galactic center'; see Elmegreen and Elmegreen \cite{elmegreen87}), a
remarkable
system of four long and regular spirals at K$'$ is reported (see Block,
Elmegreen and Wainscoat \cite{block96b} and Block \cite{block96a}).
Independent studies by Grosb{\o}l and Patsis (\cite{grosbol98}) and
Thornley (\cite{thornley96})
show that NGC 2841 is no exception: many optically flocculent galaxies
reveal a regular two-armed spiral structure in the near-infrared.

These
include
NGC 5055
and NGC 3521 (Thornley \cite{thornley96}).
Both NGC 5055 and NGC 3521 are of Elmegreen arm class 3,
but the spiral arms
of these two optically flocculent galaxies with `grand design' m=2
near-infrared counterparts
have a pitch angle at K$'$ of  13$^{\circ}$
and 14$^{\circ}$ respectively, and both galaxies are assigned
to our dust penetrated $\alpha$ bin.

If, in the midst of stellar density
waves (SDWs), SDW pressures are less than the average turbulent gas
pressure, the competition between stars and gas is won by gas: the
galaxy would present an {\it optically} fragmented, flocculent
appearance (Block \cite{block96a}). The star formation history,
principally triggered from
previous bursts of star formation (rather than SDW pressures), would
be stochastic/random. High total gas column densities (and therefore
relatively large optical depths at B or V) would be predicted.
Indeed, the V optical
depth in NGC 3223
is $\sim$ 2 at certain radii (see Block \cite{block96a}) and
it is dust which hides the m=2 near-infrared spirals.

We next turn our attention to NGC 5861 and NGC
3992. Both are described
by Sandage and Bedke (\cite{sandage94}) as having spiral patterns
which rank amongst
`the most regular in the sky'. They are exquisite grand design galaxies
which pitch angles at K$'$ of
13$^{\circ}$ and
11$^{\circ}$ respectively, and both galaxies are assigned to
our dust penetrated $\alpha$ bin.

It is therefore particularly important to note that {\it the evolved
stellar disks of some of the most exquisite grand design galaxies may
belong to the same dust penetrated class as those of the best
known optically flocculent prototypes}.
Indeed, NGC 5861 is of Elmegreen class 12 (defined in Elmegreen and
Elmegreen \cite{elmegreen87} as having `two long symmetric arms
dominating the optical disk') while NGC 5055 and NGC 3521 are of
Elmegreen arm class 3 -- almost as optically flocculent as one can
find.

NGC 4622 is an Sb galaxy which displays dual spiral structure (Byrd et
al. \cite{byrd89}). Both the outer pair and the inner arm reveal
amplitude
modulation. The inner arm is probably the first confirmed evidence
for a leading spiral arm. Based on the sequential
structure of the BVI photometry of the leading arm, beautiful evidence has
been given (Buta, Crocker and Byrd \cite{buta92}) in support of the
pattern being a counter--rotating density wave. A method to determine
the arm character (trailing or leading) in spiral galaxies using the
Fourier analysis of azimuthal profiles is presented by Puerari and
Dottori (\cite{puerari97}).

Our Fourier decomposition of the K$'$ image of NGC 4622 confirms a
strong m=1 mode which is identified with the inner (leading) arm.
The pitch angle at K$'$ for that m=1
arm is only 4$^{\circ}$.
In Block et al. (\cite{block94a}), we argued
that the leading arm
of NGC 4622 is truly exceptional and requires external driving (much like a
{\it damped} mode of a church bell), because the torques associated with
leading structures are unfavorable to self--excitation.
The rarity
of leading arms among grand design galaxies may be an indication
that tidal driving is very seldom efficient.
The pair of magnificent m=2 arms winding in the
opposite direction have a pitch angle at K$'$ of 8$^{\circ}$,
and the galaxy is assigned to our dust penetrated $\alpha$ bin.

We had earlier published a plot of
log(radius) versus azimuthal angle for our K$'$ image of NGC 2841,
where the long
arms
appear as straight lines (see colour Plate 2b in Block
\cite{block96a}).
The pitch angle of the spiral arms can easily be read off from the
slopes of the tilted features; they are found to be approximately
9$^{\circ}$. The tightly
wound arms at K$'$ of this prototypically flocculent specimen
at optical wavelengths suggests that we place it in our dust
penetrated $\alpha$ classification.

Although all the galaxies in Figures 3 and 4 show Fourier spectra where
the {\it most dominant} mode is m=2 so that they all belong to the
two-armed, evensided E$\alpha$ group,
there is nevertheless an ubiquity of the lower-order m=1 mode as well.
The only exception where m=1 is as dominant as m=2 is NGC 4622, which we
have already noted has one presumably leading arm (m=1) and a pair of
outer (presumably trailing) m=2 arms. Note, too, the
presence of higher order [m=4]
modes, but with considerably lower amplitude than the m=2 mode, in
galaxies such as NGC 3223, NGC 3521 and NGC 3992.

\section{Dust penetrated group $\beta$}

The Fourier spectra of these galaxies are presented in Figures 5 and 6.
Galaxies with a very large range in optical morphology -- from Sa to
Sc -- are to be found in this dust penetrated class.  They also span a
very wide range of Elmegreen arm class in the optical; from extreme
grand design arm class 12 to flocculent arm class 3.

We firstly turn our attention to the Sc galaxy NGC 309, of
luminosity classification I (van den Bergh \cite{bergh60}).
We have already noted that this galaxy has
some of the linearly widest Population I arms known (Block
\cite{block82}).

\begin{figure*}
     \vspace{19.5cm}
       \includegraphics{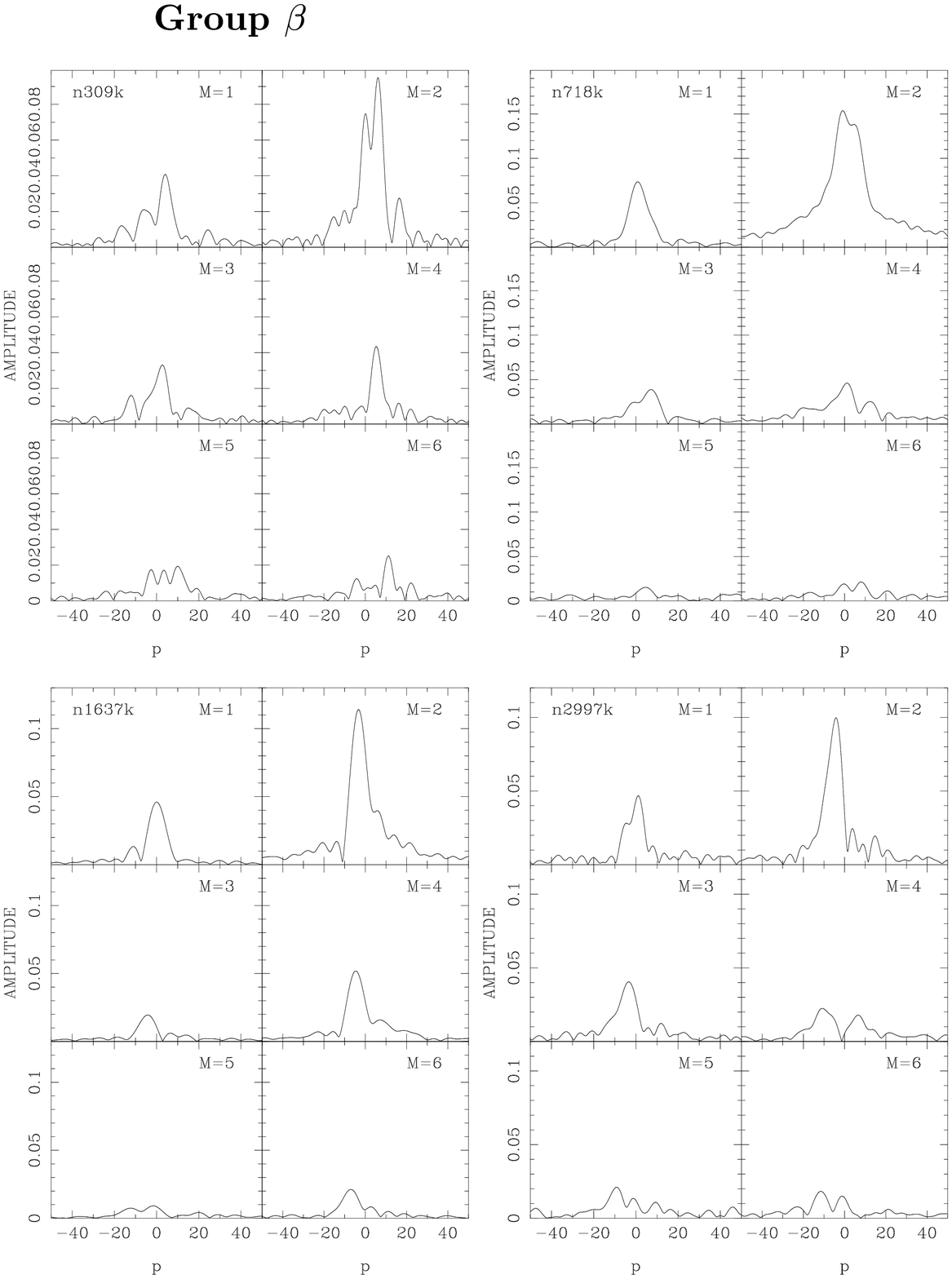}
       \caption{Fourier spectra for dust penetrated $\beta$
Population II disks. Note that although the dominant mode is m=2,
there is an ubiquity of m=1 modes as well, but with lower amplitude.
Higher order
modes (also with lower amplitude than m=2) are to be found: NGC 2997
shows a distinct m=3 component
at K$'$,  while NGC 309 and NGC 1637 reveal
significant m=4 modes in the near-infrared.} 
\end{figure*}

\begin{figure*}
     \vspace{19.5cm}
       \includegraphics{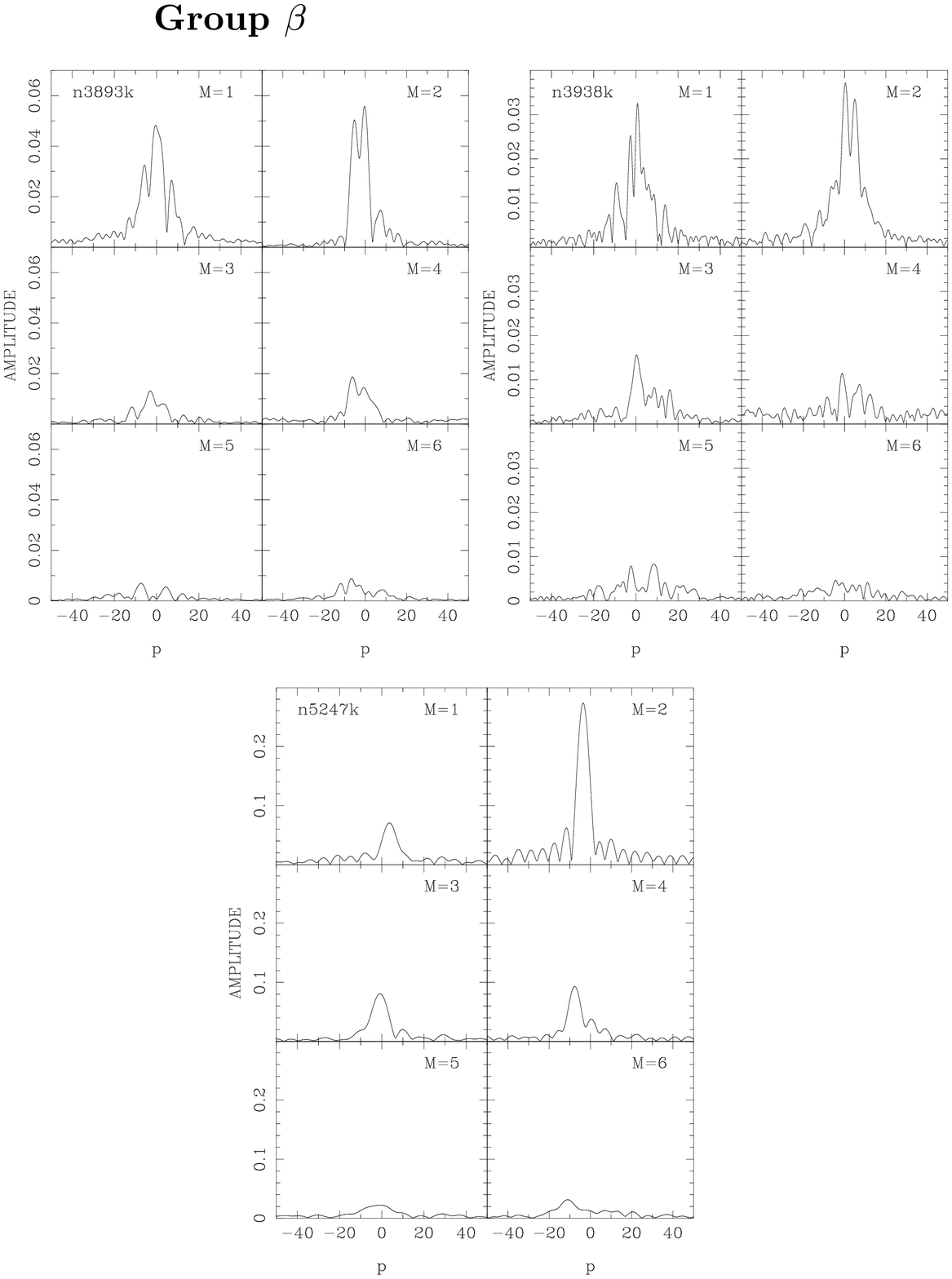}
       \caption{Fourier spectra for additional group $\beta$
class members. The dominant mode is m=2, with a manifestation of
other modes of lower amplitude. Note, for example, that apart from
m=2, NGC 5247 manifests m=1, m=3 and m=4 modes as well.}
\end{figure*}

Optically, NGC 309 presents a multi--armed morphology (see
Block and Wainscoat \cite{block91}) and the Elmegreen arm class is 9
(`two symmetric inner arms; multiple long and continuous outer arms' as
defined by Elmegreen and Elmegreen \cite{elmegreen87}).
However, at K$'$ it looks like
an SBa galaxy, betraying (1) a prominent bar and (2) two symmetrical
arms only, with winding angle just over $180^{\circ}$ and a pitch
angle of $\sim$ 18$^{\circ}$. The surface brightness
enhancements of the m=2 stellar components at K$'$ are approximately 60 \%
above that of the underlying disk.
Normal (ScI) and barred (SBa)
large scale structures co--exist within the same galaxy.

Next, we focus our attention on another galaxy whose evolved disk is
very similar to that of NGC 309: NGC 718. NGC 718 is optically
classified as SaI, but its evolved
disk belongs to the same group
at that for NGC 309. The pitch angle of its two spiral arms at K$'$ is
24$^{\circ}$ and we assign it to our dust penetrated $\beta$ bin.

Another galaxy which belongs to the same dust penetrated class as NGC
309, is the early type spiral NGC 4736. NGC 4736 is classified as type
RSab(s) and is of Elmegreen optical arm class 3. The galaxy is famous
for its inner, `knotty
ring' of HII regions (see Figure
8 in Block et al. \cite{block94a}).
Our K$'$ image reveals a pair of logarithmic
spiral arms, with a pitch angle at K$'$ of 28$^{\circ}$. The galaxy is
assigned to the dust penetrated $\beta$ bin.

Also in this dust penetrated class
is the grand--design Sc(s)I spiral NGC 2997
(Sandage and Tammann \cite{sandage87}).
Optically, the galaxy possesses two dominant spiral arms,
one of which has a prominent western branching into a
third arm (see Block et al. \cite{block94a}). NGC 2997 is of Elmegreen
arm class 9. At K$'$ the galaxy reveals a small oval distortion and a
distinct m=1 component.
The most natural
interpretation is that there are two dominant modes, m=1 and
m=2, with negative interference on the northern side.
The pitch angle of the stellar arms at K$'$ are determined to be
$\sim$ 25$^{\circ}$ and we assign NGC 2997 to our dust penetrated
$\beta$
bin.

Not only are
Sa
to Sc galaxies all to be found in this dust penetrated class, but the
dust penetrated $\beta$ bin again spans almost the entire range of
optical Elmegreen arm classes: from extreme grand design arm class 12
(eg. NGC 718) to flocculent arm class 3 (eg. NGC 4736).

From our Fourier spectra for these class $\beta$
Population II disks, we note that although the dominant mode is
m=2, there is an ubiquity of m=1 modes as well, but with lower amplitude.
The m=1 component is nearly as striking as the m=2 component in some of
the galaxies
such as NGC 3893 and NGC 3938 (Figure 6) while m=1 has a
much lower amplitude than m=2 in others, such as NGC 5247. Higher order
modes (also with lower amplitude than m=2) are also to be found: NGC
2997 shows a distinct m=3 component
at K$'$,  while NGC 309 and NGC 1637 reveal
significant m=4 modes in the near-infrared (Figure 5).

\section{Dust penetrated group $\gamma$}

The Fourier spectra for these galaxies are presented in Figure 7.

We firstly turn our attention to NGC 5195, the companion of NGC 5194 =
M51.
Optically, NGC 5195 has been classified as Ip-Ep (Morgan
\cite{morgan58}),
Irr II (Sandage \cite{sandage61}), SBa(r) (Spinrad and Harlan
\cite{spinrad72}) and SB0/a(r) (Thronson, Rubin and Ksir
\cite{thronson91}).
Our K$'$ image, radically different from the optical,
shows {\it no evidence for tidal disruption} of its old stellar population
disk by M51 (Figure 5b in Block et al. \cite{block94a}).
The
striking symmetry
in the NGC 5195 Population II disk (see the K$'$ image in Fig. 5b in
Block et
al. \cite{block94a} and confirmed by the evensided Fourier mode m=2 in
Figure 7 of this paper) is confirmed by a low Conselice number at K$'$.

\begin{figure*}
     \vspace{19.5cm}
       \includegraphics{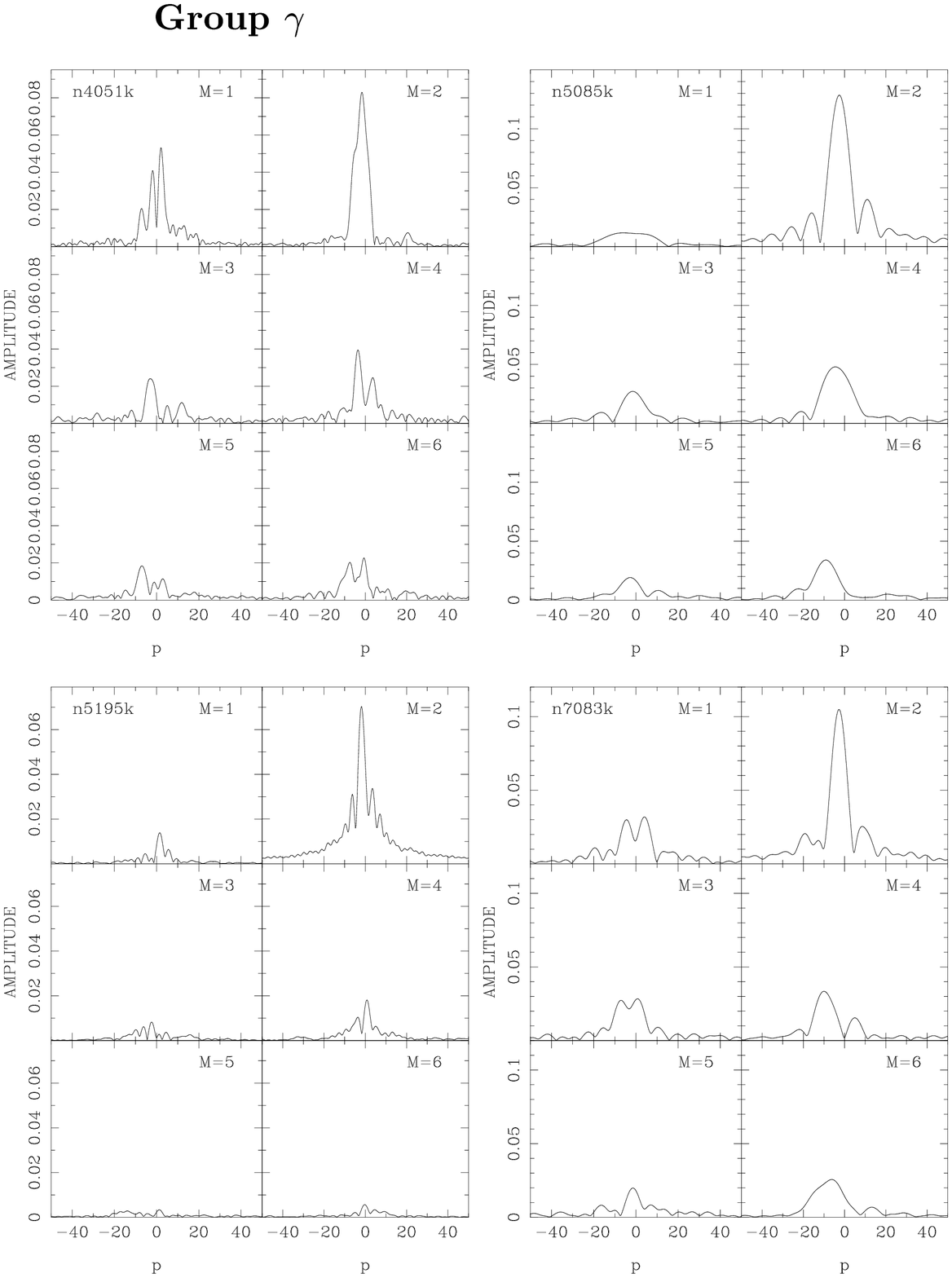}
       \caption{Spectra for dust penetrated group $\gamma$ galaxies
with open stellar arms at K$'$. The dominant Fourier mode is m=2.
Modes of higher order exist (eg. m=4 for NGC 5085).}
\label{FigGam}%
\end{figure*}

The bisymmetric structures look more like intrinsic global modes where
the m=2 spirals have a pitch angle at K$'$ of 49$^{\circ}$.
In view of the exceptional symmetry at K$'$, NGC 5195 is likely to have
{\it just disturbed} M51's spiral arms (Oort \cite{oort70}).
The tidal
interaction here is far less dramatic than usually thought, based on the
more prominent features of the lighter Population I disks.
As suggested by Block et al. (\cite{block94a}),
the prominent isophotal {\it twisting} within the bar itself may be the
quantitative signature
of the tidal field of M51 on the intrinsic bar structure of NGC 5195.
Isophotal twisting of the central bar regions is not uncommon
(Shaw et al. \cite{shaw93}) and might be simply due
to the incipient spiral structure,
commencing with a sharp turn at the bar tip, as often found for
intrinsic modes (see the discussion of NGC 1300 in Bertin
\cite{bertin93} or Fig. 10 of
Benedict et al. \cite{benedict92}, for NGC 4314).

Members of this dust penetrated $\gamma$ group include an early type
spiral (NGC 7083), a late type spiral (NGC 5085) as well as an optical
irregular. NGC 5085 is of Elmegreen arm class 2 (`fragmented spiral arm
pieces with no regular pattern'). Also included is NGC 4051, of
Elmegreen
arm class 5 (`two symmetric short arms in the inner regions; irregular
outer arms').

The Fourier spectra for dust penetrated group $\gamma$ galaxies
with open stellar arms in Figure 7 show that the dominant Fourier
mode is m=2. However, modes of higher order do exist (eg. m=4 for NGC
5085).

\section{Classifying lopsided galaxies}

NGC 1637, for example,
draws attention
because of its exceptional lopsidedness, with a
somewhat squarish shape, although there is a regular bisymmetric
spiral close to the nucleus. It is classified as Sc (Sandage 1961),
but shows a marked
lopsidedness (an m=1 feature) in the old Population II disk (see Figure
9b in Block et al. \cite{block94a}).

If one wishes to draw attention to lopsidedness
in galaxies displaying a prominent m=1 feature,
we propose that
the galaxy still be placed in the $\alpha$, $\beta$ or $\gamma$ bin,
according to the pitch angle of the m=1 component, but that a prefix
L clearly be used to designate the lopsidedness in the evolved disk.
The m=1 feature itself may be prominent, but not dominant, as in NGC
1637.

We have carefully examined the m=1 feature at K$'$ of NGC 1637 in a
$({\rm ln} R,
\theta)$ diagram and find a straight line, indicating that the fit of a
logarithmic spiral to that m=1 arm is excellent. From the slope, we
find the pitch angle of the
m=1 arm to be $\sim$ 35$^{\circ}$.

We have analysed the inner
bisymmetric spiral by setting an appropriate annulus to emcompass
the inner spiral but exclude the outer m=1 arm.
The spiral shows up clearly at K$'$ in the spectra, and the
pitch angle of the dominant m=2 component (Figure 5) is 29$^{\circ}$.
For the regular m=2 bisymmetric spiral, a dust penetrated E$\beta$ would
be appropriate.

Binning according to the
asymmetric/lopsided mode of m=1 here, the m=1 spiral belongs
to the dust penetrated $\gamma$ bin  and would be designated
L$\gamma$. The pitch angle of the lopsided m=1 arm is 6$^{\circ}$
greater than for the m=2 mode. In Figure 1, NGC 1637 is
included
by plotting an m=1 logarithmic spiral with a pitch angle of 35$^{\circ}$.

No companion for NGC 1637 is found on
the Palomar Observatory Sky Survey within one degree. Such extreme
lopsidedness could be a general feature in some
isolated galaxies, for which the persistence of such m=1
asymmetries (Baldwin, Lynden-Bell and Sancisi \cite{baldwin80})
should be
explained. As exemplified by these authors, lopsided prototype galaxies
such as NGC 891 and 5457, are indeed
lopsided both in the light and in the
neutral HI gas. Within the modal theory,
one may argue that lopsidedness and one--armed structures
are all non-linear m=1 modes. In the linear theory they
may be the m=1 equivalent of what is more commonly noted as barred and
normal two--armed spiral modes (the former occurring in heavy, fully
self--gravitating disks and the latter occurring instead in
light disks). If this is so, such an extreme lopsidedness
in the stellar disk would be more natural if the disk is heavy. Indeed, a
{\it prominent bar} is observed in the old stellar disk of NGC 1637.

\section{Higher order harmonics}

While there is an ubiquity of m=1 and m=2 modes in our
sample (Figures 3--7), higher order modes in some instances do
exist, as we
have already pointed out. However, in no galaxies in our sample are the
higher order modes the {\it dominant} ones.

In the K-band study of 45 spirals by Seigar and James
(\cite{seigar98b}), the most common
dominant modes were also found to be the lopsided L (m=1) and evensided
E (m=2) --  each of
which was found to arise in  their sample in about 1/3 of
the galaxies.

The sample of Seigar and James (\cite{seigar98b}) showed a great dearth
of spirals with dominant m=3 (only 4 out of the 45 cases), but
nevertheless
such galaxies do exist and one must be able to classify them in a
dust penetrated regime.
Seigar and James (\cite{seigar98b}) did find that approximately 25\%
of their total showed a dominant m=4 mode in the K-band.

We would suggest the terminology H3 and H4
for the third and fourth harmonics, to assist with
easy visualisation of their evolved disk morphologies.

Can modes greater than m=2 develop within the modal theory of spiral
structure?
Already in the paper by Block et al. (\cite{block94a}) and by
Bertin (\cite{bertin96b}) it was hinted that:

Firstly, in a {\it gas-rich} system some higher-m modes should develop,
and that might
also induce some response in the stellar disk, for the stronger cases.

Secondly, non-linearly modes may couple and again give rise to
higher-m structures 2 + 1 = 3 and 2 + 2 = 4 (Block et al.
\cite{block94a} and Bertin, private communication). A variation on
this theme is the following: if one takes a strong bisymmetric bar and
Fourier-analyze the deprojected image, one may indeed find m=4 and
even m=6 to be prominent, even
when one does not see actually four or six arms. If the bar is
strong
enough to drive the disk, an m=4 response in the outer disk may show
up. It is interesting to note that almost all of the m=4 cases cited by
Seigar and James (\cite{seigar98b}) are classified as barred in the optical.

Thirdly: another point to consider for the generation of higher-m
modes is the underlying rotation curve.
If the shear is {\it very} mild (some Sc's have a large part of their
disks in
almost solid body rotation), then the ILR is less effective (Bertin,
private communication) and m=3 and even m=4 modes might develop.

We suggest that these H3 and H4 galaxies, too, should
be binned in the dust penetrated regime, according
to the pitch angle of their stellar arms.

All of the H3 and H4 galaxies in Seigar and James (\cite{seigar98b})
belong
to our dust penetrated $\alpha$ class, so that a galaxy such as UGC
2303 (m=3, with a pitch angle at K of 4$^{\circ}$) would be classified
here as H3$\alpha$, while a spiral such as IC 357 (m=4, with a K pitch
angle of 8$^{\circ}$) is classified here as H4$\alpha$.

\section{Inverse Fourier Transform analysis}

\begin{figure*}
     \vspace{18.7cm}
    \caption{
Contours of the inverse Fourier transform are overlaid on
deprojected 2.1$\mu m$ images of NGC 3223 (Hubble type b: top
left) and NGC 5861 (Hubble type c; top right)
to indicate the tight winding of
the
stellar arms of the $\alpha$ class. The pitch angle of the K$'$ arms
becomes more open for the $\beta$ bin: illustrated are NGC 718 (Hubble
type a; middle left)
and NGC 309 (Hubble type c; middle right).
Finally, the $\gamma$ bin includes
NGC 5085
(Hubble type c; lower left) and NGC 7083 (Hubble type b; lower right) with
wide open arms in their evolved
disks. There is no correlation between the dust penetrated archetypes and
optical Hubble type. }
\end{figure*}

In Figure 8,  we present the inverse Fourier transforms of six of the
galaxies in Table 1, two from each $\alpha$, $\beta$ and $\gamma$ dust
penetrated classes.

After having
deprojected the K$'$ images and identifying the dominant modes, we next
calculated the inverse Fourier transform, as follows:

We define
the variable $\displaystyle u \equiv
{\rm
ln}\;r$. Then

$$\displaystyle S(u,\theta) = \sum_m
S_m (u) {\rm e}^{im \theta} $$

where

$$\displaystyle S_m(u) = \frac{D}{{\rm e}^{2u} 4 \pi^2} \int_{-\infty}^
{+\infty}\;G_m (p) A(p,m) {\rm e}^{i p u}\;dp$$

and

$$\displaystyle D=\Sigma_{i=1}^I \Sigma_{j=1}^J I_{ij} (u,
\theta)$$

$G_m(p)$ is a high frequency filter used by Puerari and
Dottori (\cite{puerari92}). For the spiral with

 $\displaystyle \tan P = -\frac{m}{p_{max}^m}$ it has the form

$$\displaystyle G_m(p) = {\rm exp} \left[ -\frac{1}{2} \left( \frac{p
- p_{max}^m}{25} \right)^2 \right] $$

\noindent where $p_{\rm max}^m$ is the value of $p$ for which the
amplitude of
the Fourier coefficients for a given $m$ is maximum. This filter is also
used to smooth the $A(p,m)$ spectra at the interval ends (see Puerari
and Dottori \cite{puerari92}).

The contour overlays of the inverse Fourier transforms in
Figure 8 indicate
the excellent fit of our pitch angles, robustly derived from the
Fourier spectra. In the top panel of Figure 8, two galaxies with
tightly wound pitch angles at K$'$, belonging to the $\alpha$ class,
are presented. The galaxies are NGC 3223 (optical Hubble type b) and NGC
5861 (optical Hubble type c). The fit of logarithmic spirals to the
arms is excellent (Figure 8), as was already alluded to by the
pioneering empirical rectifications of logarithmic spirals by Danver
(\cite{danver42}).
In the middle panel of Figure 8, two galaxies (NGC 718 and NGC 309) are
presented; they belong to class $\beta$; again the overlay of the
inverse Fourier transform confirms the robustness of the Fourier
technique of identifying the dominant mode and binning according to
pitch angle at K$'$. Note that NGC 718 is of Hubble type a, while
NGC 309 is of Hubble type c.
Finally, contour overlays of the Fourier
transform for two of the class $\gamma$ spirals (NGC 5085 -- Hubble
type c and NGC 7083 -- Hubble type b) are illustrated in the lower panel
of Figure 8.

\section{Discussion}

The question of whether star formation actually traces underlying
spiral structure in the Population II disk -- in other words, whether
spiral density waves actually trigger star formation -- has been
elegantly
reviewed by Elmegreen (\cite{elmegreen95}). The issue is whether the
interstellar
gas does not form many stars while it is in the interarm regions, but
rapidly forms stars when it enters a stellar spiral arm.

Since an optically flocculent galaxy can belong to the same dust
penetrated bin as that of an optically grand design and both have
a dominant m=2 mode in the near-infrared,
the implication is that stellar density waves do not
principally  trigger
star formation -- for if they always did, the morphology of the
optical
disk would closely mimic its dust penetrated archetype. Only if star
formation
traced density waves could one {\it predict} what the underlying mass
distribution of a galaxy would be and what the dust penetrated class
would be, which one cannot. NGC 3521 has a regular two-armed stellar
mass distribution; in the optical, this
galaxy is flocculent.

Important confirmations of this result are to be found in the
literature.
Elmegreen and Elmegreen (\cite{elmegreen86}) showed that the star
formation rate per unit area, from UV, FIR, H$\alpha$, integrated
colour, and surface brightness data, is the same for grand design
and flocculent galaxies.
A study by McCall and Schmidt (\cite{mccall86}) showed that
the fraction
of galaxies with grand design spirals is the same whether they have
Type I, Type II, or indeed no recorded supernovae, which implies that
this fraction is independent of the supernova rate.

Similar inferences have been made for our Galaxy. Talbot
(\cite{talbot80}) found that the star formation rate per unit molecular
mass in the Milky Way is independent of radius, and
Wouterloot, Brand and Henkel (\cite{wouterloot88}) found the same star formation
rate per unit
molecular mass for molecular clouds in the far-outer disk of our Galaxy.

Viewed at arcsecond resolution, it has become clear that interarm dust
(and gas)
can be very widespread, covering much of the optical disk of a spiral
galaxy (eg. Block \cite{block96a}).
Giant molecular associations are to be found in flocculent as well as
grand design spirals (eg. Sakamoto \cite{sakamoto96}).
A unified view of macromolecules, very small grains and large dust
particles in the Whirlpool Galaxy M51
has recently been produced by combining optical and Infrared
Space Observatory data (see Block
et al. \cite{block97}). Widespread spirals of macromolecules
and dust grains
lie not
only in the arm, but also in the interarm regions (see Figure 1 in
Block et al. \cite{block97}).
If these spirals are indeed precursors to density wave triggered star
formation, a {\it partial} decoupling of the Population I and II disk is
suggested (the reader is referred to the `smooth red arms' and `bar' in
the Population II disk of M51 shown by the photographic work of Zwicky
\cite{zwicky57}).

Finally, there are indications from rotation curve analyses that
optical Hubble type is not correlated with the evolved Population II morphology.
Indications are that the distribution
of stellar mass in the disk, to which the properties of rotation
curves are tied, can be different for the {\it same}
Hubble type (see Burstein and Rubin \cite{burstein85}).
These authors find three principal types of mass distribution, with
Hubble type a and b classes being found {\it among all three types}
more or less equally. To secure rotation curves, one selects
galaxies from a sample which are well inclined to the line-of-sight;
whereas for morphology, a
preferentially face-on criterion is essential to delineate and study
the arms.
If the arms could clearly be delineated in their sample of (inclined)
galaxies, it would not
be unreasonable to predict that a correlation might be found between our
dust penetrated classes and their three principal types of mass
distribution.

\section{Conclusions}

To derive a coherent physical framework for the excitation
of
spiral structure in galaxies, one must consider the co-existence of
{\it two different} dynamical components: a gas-dominated Population
I disk (OB associations, HII regions, cold interstellar HI gas) and an
evolved stellar Population II component. The Hubble classification
scheme has as its focus, the morphology of the Population I component
only.

Our first major conclusion of this observational study is that
there is an ubiquity of
m=1 and m=2 modes in the near-infrared regime (Figures 3--7).

Secondly,
three principal
archetypes $\alpha$, $\beta$ and $\gamma$ for the dust penetrated
morphology may be
proposed, characterised by pitch angles at K$'$ of $\sim$
10$^{\circ}$ (the $\alpha$ class), $\sim$ 25$^{\circ}$ (the $\beta$
class) and $\sim$ 40$^{\circ}$ (the $\gamma$ bin), respectively (see
Figure 1).

The
ubiquity of low-m features finds a natural interpretation within
the modal theory, because higher-m modes are expected to be damped by
absorption at the ILR in the evolved stellar
disk. If the shear in a galactic disk is not high, it is possible
for higher-m modes to develop as the ILR is then not as effective. Such
higher-m modes may also develop for gas-rich galactic disks.

Thirdly, the dynamical behaviour of gas and young OB stars is
often {\it decoupled} or partially decoupled from that of Population II
disk, and this explains why an Sc can mimic an SBa at 2.1$\mu m$.

\begin{figure}
     \vspace{7.5cm}
       \includegraphics{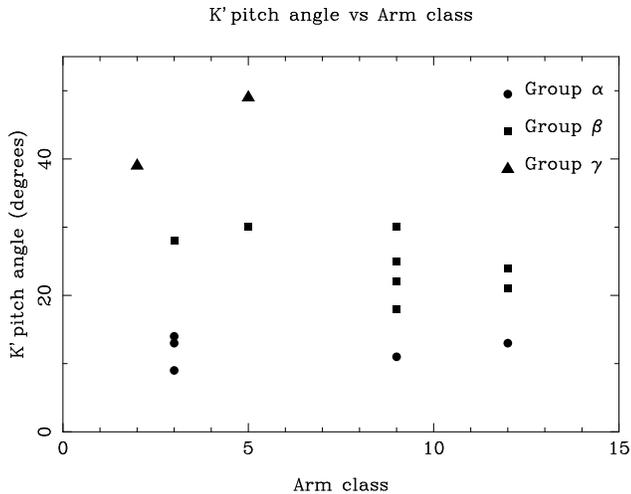}
       \caption{Pitch angle at K$'$ {\it vs.} optical Elmegreen arm
class.
Optically flocculent galaxies of arm class 3 can have almost
identical evolved disk morphologies to those belonging to optically
grand design spirals of arm class 12.}
\end{figure}

\begin{figure}
     \vspace{7.5cm}
       \includegraphics{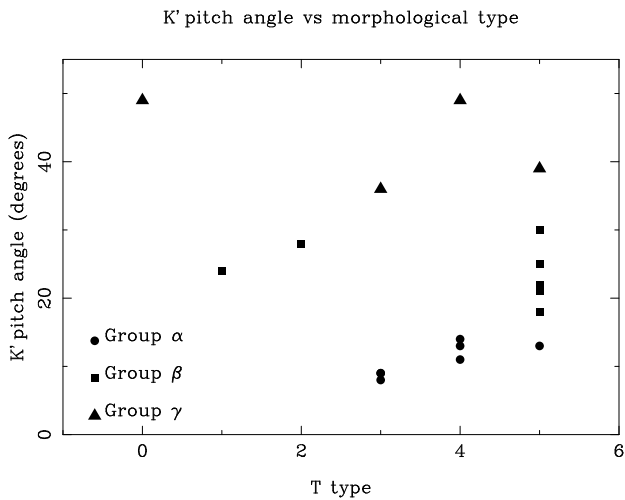}
       \caption{Pitch angle at K$'$ {\it vs.} Morphological type -- de
Vaucouleurs' T index. The  spiral structure in the dust penetrated
disk of NGC 5195 (T=0) can have the same pitch angle as that for NGC
4051 (T=4).}
\end{figure}

Fourthly, and related to the decoupling: we find no correlation
between optical Elmegreen arm class
and pitch angle in the dust penetrated regime (see Figure 9)\footnote{A short
aside: Four of the galaxies in our sample do not have listed Elmegreen
arm
classes, as they were not included in the
sample of Elmegreen and Elmegreen (\cite{elmegreen87}).}.
In our $\alpha$
bin, for example, NGC 3521
is of Elmegreen class 3, NGC 5861 is of Elmegreen class 12, yet both
have pitch angles differing by only one degree at K$'$ (see Table 1). In
the $\beta$ bin, the variation in Elmegreen arm class in Table 1 is
again from flocculent (3) to grand design (12).
Figure 9 clearly shows why a completely new classification
of disks in the near-infrared domain is called for:
two galaxies having completely different Elmegreen arm classes can have
almost identical pitch angles at K$'$. In Figure 10, we plot a
similar
graph, but of pitch angle at K$'$ versus de Vaucouleurs' T index. The
$\alpha$
dust penetrated archetype can be the resident disk to a vast range of
optical Hubble or de Vaucouleurs' T indices. Likewise for the
$\beta$ and $\gamma$ classes.

Although the m=2 mode is a dominant mode in our Fourier spectra
(Figs. 3--7),
there is often a significant m=1 component as well.
If one wishes to draw
attention to the m=1 mode,
we have proposed that a prefix L indicate the lopsidedness; the dust
penetrated classes would then be L$\alpha$, L$\beta$ and L$\gamma$.
Any of the L archetypes will look like one
logarithmic spiral (instead of two) in Figure 1, determined by the
appropriate pitch angle. In that Figure, we have plotted one
logarithmic spiral for the m=1 mode of NGC 1637 and for the (leading)
arm in NGC 4622.
We use the terminology E=evensided for dominant two-armed (m=2) modes,
and H3 and H4 for third and fourth order harmonics.

It may also be appropriate, in due course, to introduce a class
F=fragmentary, for stellar disks which are patchy
and irregular.
It is important that the letter I=irregular not be used, since a few
optically flocculent and late type Sd {\it spiral} galaxies, for
example, can belong to this group.
An example of an F class would be the flocculent type Sd companion
to M81,
NGC 2976 (see section 2.3 in Block \cite{block96a}), which presents a very
irregular appearance in the near-infrared -- no central mass concentration
and no coherent stellar morphology at all.

As in the optical scenario,
it cannot be expected of any classification scheme to include {\it every
single} galaxy. The cosmic tapestry, especially at high z, is too rich
and varied. There is one additional crucial point pertaining to galaxy
morphology classification in the high redshift Universe: the
detection
of an `old' stellar disk (ages greater than 1 or 2 Gyr) beyond a
redshift of 1.5 is extraordinarily difficult -- see Figure 6 in Abraham
(\cite{abraham98}).
As one probes progressively deeper into redshift space, the rest frame
of a galaxy systematically shifts to the ultraviolet. This is the
morphological
K-correction: one is observing galaxies at bluer rest wavelengths as a
function of redshift. Strong surface brightness selection effects
bias against the detection of even intermediate-aged stellar
populations (see Abraham \cite{abraham98}) and the decline in visibility
of older stellar
populations with redshift is dramatic.  The
late type/`irregulars' reported in many of the higher redshift
galaxies in the Hubble Deep Field (HDF) describe {\it ultraviolet}
morphologies of the {\it gaseous} disks only. Some of these disks {\it
could possibly be} partially
or fully decoupled from their stellar disks as we find locally for the
optical
irregular in the system NGC 5195/M51. (A promising note is that Fourier
decompositions can be done on HDF images. This is provided
that the number of components used in the decomposition is not too large
(Windhorst, private communication) since faint galaxy images are
so small, even with the Hubble Space Telescope).

Returning to the F class: while some optical irregulars may belong
to this type,
the dust penetrated disks of other optical irregulars (eg. NGC 5195) would
not.
We believe that while modes
have been identified in large ensembles of dust penetrated galactic
disks in our local low redshift Universe, there ofcourse will always be
exceptions
where there are no indications of any spiral arms or of  modes -- NGC
2976 is one example.

We have deliberately kept our preliminary dust penetrated classification
scheme {\it simple}. Of course it is possible to
use the terminology (B)E$\alpha$, (B)E$\beta$ and (B)E$\gamma$ for
example, to indicate the
presence of a bar in the evolved disk of evensided galaxies
-- and possibly to give a measure of bar strength in terms of the
`equivalent angle' of Seigar and James (\cite{seigar98a}, see their
Figure 10). However, since it is not uncommon for a barred and unbarred
morphology
to co-exist within the same galaxy, we have only differentiated evolved
disks here on the basis of the dominant Fourier mode
and on the opening or pitch angle of the stellar arms.
It is a starting point. The caution of Mike Disney (\cite{disney96})
rings in our ears:

{\it ``What would worry me is that everybody will go off now and write
their own morphology and we'll have 500 more different K-band
morphological categories. I don't know how to solve it; perhaps several
people at this conference should go away and agree to do it. At least to
have some starting places for it -- otherwise it may be as haphazard as
it's been in the past in the optical.''}

In this paper, we have tried to stress the {\it
fundamental} rather than {\it incidental}
need to develop a near-infrared classification scheme of
galaxies. Up to
now, we have only classified one of the two components: the Population I
disk.  A central aspect here is the likely coupling
of the Population I disk with that of the Population II disk via
a {\it feedback} mechanism.

Having formulated our dust
penetrated classification
scheme above, we have tested it on an independent sample of 45
face-on galaxies observed in the K-band by Seigar and James
(\cite{seigar98b}). It is interesting to note that
Seigar and James did not
study galaxies with very open arm morphologies in the
near-infrared. Indeed,
44 out of their 45 galaxies have pitch angles at K less than
13$^{\circ}$, which designate them to our $\alpha$ bin.
Nevertheless, this strengthens our conclusion even further, that {\it
one} specific
dust penetrated archetype (in this instance, class $\alpha$) may be the
resident disk to a further 44 spirals with a wide range in optical
morphologies: from early (a) to late type (d) in their Table 1.

It is perhaps important to reflect back to the thoughts of Hubble
(\cite{hubble36})
in deciding which features should be included, and which features should
be excluded, when classifying galaxies:

{\it ``The features must be significant -- they must indicate physical
properties of the nebulae [galaxies] themselves and not chance effects
of orientation -- and also they must be conspicuous enough to be seen in
large numbers of nebulae.''}

We see a duality of spiral structure. One for the Population I disk;
often a radically different one for the Population II disk. One half of
the story has been missing. A feedback between a young Population I
disk
of an optically flocculent galaxy, and an old grand design
Population II disk, for example.

Finally, the morphology of evolved disks with open stellar arms
could be interpreted as favouring the secular evolution of galaxies --
from an open to a more tightly wound scenario: the
redistribution of angular momentum by large-scale spiral torques
being more efficient for those stellar arms which are more open
(eg. Pfenniger et al. \cite{pfenniger96}; Combes and Sellwood,
private communications).

\begin{acknowledgements}

The deep insight of our referee, Dr P. James, from both an observational
and theoretical point of view, has been especially helpful.
DLB is deeply grateful for invitations to be a Visiting Astronomer
at the Institute of Astronomy in Hawaii and at the European Southern
Observatory (Garching and Chile) where much of the work was
initiated. It is a great pleasure to have observed at Mauna Kea
with Dr A. Stockton, and for his unfailing
input of K$'$ images over the years.
The present Fourier study was undertaken while one of us (IP) was a
Distinguished
Visitor in the Department of Computational and Applied Mathematics at
the University of the Witwatersrand. The Distinguished Visitor's Program
is funded by the Anglo-American and de Beers Chairman's Fund Educational
Trust to which IP and DLB wish to record a special note of gratitude.
We thank
Drs. P. Grosb{\o}l, M. Verheijen and M. Thornley for sharing their K$'$
images with us. DLB is deeply grateful to Giuseppe Bertin, Bruce
Elmegreen, Fran\c coise Combes, Jerry Sellwood and Rogier Windhorst for
important
discussions.

\end{acknowledgements}

\end{document}